\documentclass[aps,prl,reprint,groupedaddress,shownopacs,amsmath,amssymb,twocolumn,floatfix]{revtex4}
\usepackage{txfonts}
\usepackage{mathrsfs,times}
\usepackage{ulem}
\usepackage{textcomp}

\usepackage{graphicx}
\usepackage{epstopdf}

\usepackage[applemac]{inputenc}
\usepackage[T1]{fontenc}
\usepackage[english]{babel}
\usepackage{ae}
\usepackage{siunitx}
\usepackage{color}
\usepackage{url}

\usepackage{amsmath,amssymb,natbib}
\usepackage{psfrag}
\usepackage{fixmath}
\usepackage{booktabs}

\usepackage{slashed}

\usepackage[americaninductors]{circuitikz}
\usepackage{tikz}
\usetikzlibrary{arrows}

\begin{document}

\title {Scalable Ion Trap Architecture for Universal Quantum Computation by Collisions }

\author{Pengfei Liang$^{1,2}$}
\author{Lingzhen Guo$^{1,3}$}
\email{lingzhen.guo@mpl.mpg.de}
\affiliation{$^1$\mbox{Institut f\"ur Theoretische Festk\"orperphysik (TFP), Karlsruhe Institute of Technology (KIT), 76128 Karlsruhe, Germany}\\
$^2$\mbox{Department of Physics, Beijing Normal University, Beijing 100875, China}\\
$^3$\mbox{Max Planck Institute for the Science of Light, Staudtstr. 2, 91058 Erlangen, Germany}\\}

\date{\today}

\begin{abstract}
We propose a scalable ion trap architecture for universal quantum computation, which is composed of an array of ion traps with one ion confined in each trap. The neighboring traps are designed capable of merging into one single trap. The universal two-qubit $\sqrt{SWAP}$ gate is realized by direct collision of two neighboring ions in the merged trap, which induces an effective spin-spin interaction between two ions. We find that
the collision-induced spin-spin interaction decreases with the third power of two ions' trapping distance. Even with a \SI{200}{\micro\meter} trapping distance between atomic ions in Paul traps, it is still possible to realize a two-qubit gate operation with speed in \SI{0.1}{\kilo\hertz} regime.
The speed can be further increased up into \SI{0.1}{\mega\hertz} regime using electrons with \SI{10}{\milli\meter} trapping distance in Penning traps.
\end{abstract}

%\pacs{37.10.Ty, 03.67.Lx, 03.75.Lm, 71.70.Gm}

\maketitle

The hyperfine electronic levels of an ion usually have very long coherence time \cite{Nature-476-2011,PRL-113-2014,NaturePhotonics-11-2017}, which makes the ion trap architecture a very promising platform for universal quantum computation \cite{Nature-453-2008,Nature-464-2010}.
By applying laser beams with proper frequencies to each ion, one can initialize qubit state \cite{JMO-54-2007,PRL-113-2014}, operate single-qubit gate \cite{PRL-113-2014} and readout qubit state \cite{OL-38-2013,PRL-113-2014} with high efficiency. The challenge comes from coupling two ionic qubits and realizing fast universal two-qubit gates \cite{PRL-75-1995,NJP-113060-2015}. The conventional solutions include mediating the interaction between two ionic qubits by phonons\cite{PRL-74-1995,PRL-82-1999,PRL-91-2003} or phtotons \cite{Nature-402-1999,PRL-78-1997}. However, it is still an outstanding challenge to scale the trapped ions to enough numbers to outperform the classical processors \cite{Nature-404-2000,Nature-417-2002,Science-339-2013,SciAdv-3-2017}, and scale the ability to individually address the qubits in the chain of ions \cite{APL-97-2010,Nature-471-2011}. In this letter, we proposed a new ion trap architecture for universal quantum computation with advantages of both scaling large number of ions and addressing individual ions. Different from mediating interaction via a common bus, we couple the two ionic qubits via direct collision of two ions in the trap.

\begin{figure*}
\centerline{\includegraphics[height=0.2\textheight]{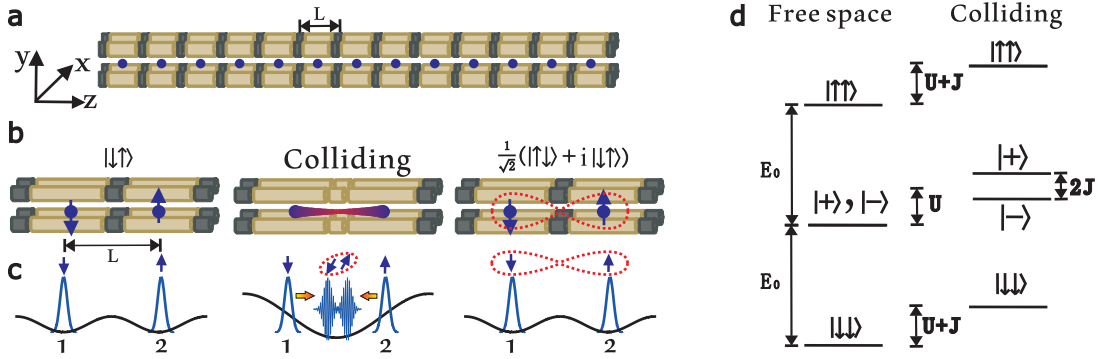}}
\caption{\label{fig_CIT}{\bf{Ion trap architecture for universal quantum computation via collision-induced spin-spin interaction.}} {\bf a}, the basic structure of our ion trap architecture, which is an array of ion traps with length $L$. The long rods (light colors) represent the quadrupole electrodes of linear Paul traps connected by link electrodes (dark colors). {\bf b}, three processes to realize the universal two-qubit $\sqrt{SWAP}$ gate. The left figure shows the spin product state $\left|\downarrow\uparrow\right\rangle$ of two ions trapped in two separated traps. The middle figure shows the two traps emerge into one single trap and the two ions are colliding with each other. The right figure shows the single trap are split into two traps again and the two ions are confined in the two separated traps when the target spin entangled state (indicated by red dashed tangled circles) is arrived. {\bf c}, the confinement potentials in the longitudinal $z$ direction and the quantum states of two ions, i.e., real parts of wave functions in Eq.~(\ref{phi}), corresponding to the processes shown in figure b. The left figure shows two ground states of ions confined in two separated traps. The middle figure sketches the colliding process of two ions in the merged large trap. The red dashed circle indicates the spins of two ions start to interact and entangle with each other during colliding. The right figure shows the two ions are confined on the bottoms of two separated traps again with spins entangled (indicated by red dashed tangled circles). {\bf d}, the internal energy level structures of two ions in free space without Coulomb interaction (left) and in the single merged trap during colliding (right). The collision induces a common level shift $U$ and a level splitting $2J$ between the triplet state $\left|+\right\rangle$ and singlet state $\left|-\right\rangle$. $E_0$ is the hyperfine level splitting single ion.
}
\end{figure*}

In Fig.~\ref{fig_CIT}a, we sketch the envisioned architecture of ion traps for universal quantum computation. The unit component of our architecture is the linear ion trap (Paul trap) with length $L$, which is basically composed of quadrupole electrode (long rods with light color) plugged at two ends with short segmented electrodes (rods with dark color) \cite{RMP-531-1990,JRNIST-103-1998}, which we call the {\it link electrodes} here for convenience. A radio frequency (rf) potential is applied between diagonally opposite rods in the quadrupole electrode giving rise to (harmonic) ponderomotive potentials in the transverse $x$ and $y$ directions with confinement frequencies $\omega_x$ and $\omega_y$ respectively \cite{JRNIST-103-1998}. By applying additional electric potentials to the link electrodes beside the quadrupole electrode, a confinement potential with characteristic frequency $\omega_z$ in the longitudinal $z$ direction is provided. Usually, the transverse confinement is symmetric ($\omega_x=\omega_y\equiv\omega_{x,y}$) and much stronger than the longitudinal confinement, i.e., $\omega_{x,y}\gg\omega_z$. In this work, we define the confinement parameter $\omega_\bot\equiv\omega_{x,y}/\omega_z$ to describe the level of transverse confinement. In our architecture, each trap confines only one ion, which is convenient for individual-qubit addressing. The ions are cooled on the bottoms of traps by Doppler laser cooling technique \cite{RMP-75-2003}. Two stable hyperfine levels of ions, e.g., the ground-state $^2S_{1/2}\left|F = 0, m_F = 0\right\rangle$ and $^2S_{1/2}\left|F = 1, m_F = 0\right\rangle$ of ion $^{171}\mathrm{Yb}^+$ with hyperfine level splitting $E_0/\hbar\approx 2\pi\times \SI{12.64}{\giga\hertz}$, are utilized to serve as the qubit states characterized by an effective spin-up state $\left|\uparrow\right\rangle$ and spin-down state $\left|\downarrow\right\rangle$ corresponding to bit values $1$ and $0$ respectively \cite{PRA-013406-2011}. In principle, the linear Paul traps for atomic ions in our architecture can be replaced by (open-endcap) Penning traps for electrons \cite{PS-34-47-1991,IJMSIP-88-1989}, where the transverse confinement is applied by a strong homogeneous magnetic field along the longitudinal direction and the qubit states are the real spins of electrons.

In Fig.~\ref{fig_CIT}b, we sketch the basic three processes to couple the two ions and entangle their spins by direct collisions. In the left figure, two ions are confined on the bottoms of two neighboring traps by applying static electric potentials on the three link electrodes. The internal qubit states of two ions are initialized to a product state with one spin down and the other spin up, i.e., $\left|\downarrow\uparrow\right\rangle$. In the middle figure, we merge the two neighboring traps into one single large linear trap by synchronizing the potential on the middle link electrode with the rf potentials on the quadrupole electrodes beside. Since the two ions are no longer on the bottom of the merged trap, they start to oscillate in the trap and collide with each other. The key mechanism used our proposal is that the collision induces an effective spin-spin interaction between two ions \cite{arXiv-1710-09716}, which we call {\it collision spin effect}. As a result, the spins of two ions start to rotate and entangle with each other during the collision. In the right figure, we split the single ion trap into two traps by tuning the potential on the middle link electrode back to the initial condition and confine the two ions on the bottom of each trap again. By controlling the collision time or the number of collisions, it is possible to arrive at a spin entangled pair such as $\frac{1}{\sqrt{2}}(\left|\uparrow\downarrow\right\rangle+i\left|\downarrow\uparrow\right\rangle)$, and realize the universal two-qubit gate $\sqrt{SWAP}$ gate. More complicated gate operations for quantum computation can be operated by merging and splitting neighboring traps sequentially in the architecture.

As discussed above, the key process is the collision of two ions in the trap. We shall discuss in detail the two-ion dynamics during collision. In the semi-classical limit and neglecting the Coulomb interaction between ions, the dynamics of two ions in the merged single ion trap are described by two oscillating coherent states. In the coordinate representation, the wave functions of two coherent states are given by
\begin{eqnarray}\label{phi}
\left\{
\begin{array}{lll}
\phi(z_1,t)=N(t)\exp{\left(-\frac{1}{2z_0^2}\big[z_1+x(t)\big]^2+i\frac{p(t)}{\hbar}z_1\right)}\\
\varphi(z_2,t)=N(t)\exp{\left(-\frac{1}{2z_0^2}\big[z_2-x(t)\big]^2-i\frac{p(t)}{\hbar}z_2\right)}.\\
\end{array}
\right.
\end{eqnarray}
Here, $x(t)\equiv\frac{1}{2}L\cos(\omega_zt)$ and $p(t)\equiv\frac{1}{2}m\omega_zL\sin(\omega_zt)$ are the position and momentum parameters of two ions in the trap. $N(t)\equiv\pi^{-\frac{1}{4}} z_0^{-\frac{1}{2}}\exp{\big[i\frac{L^2\sin(2\omega_zt)}{16z_0^2}}\big]$ is the common normalization factor of two coherent states with $z_0\equiv\sqrt{{\hbar}/{m\omega_z}}$ the characteristic length of coherent state and $m$ the mass of single ion. We see that the two oscillating coherent states can be viewed as plane waves with instantaneous wave vectors $\pm p(t)/\hbar$ and Gaussian envelop centered at $\mp x(t)$ with characteristic width $z_0$ .
When we merge the two separated traps by the middle link electrode, we should tune the potentials on the two side link electrodes correspondingly to keep the longitudinal trapping frequency $\omega_z$ unchanged. In the quantum dynamics, due to the indistinguishability, we should reconstruct the normalized symmetric and antisymmetric spatial wave functions of two ions by
\begin{eqnarray}\label{sym}
\left\{
\begin{array}{lll}
\Phi^+(t)&=&\frac{1}{\sqrt{2\big(1+e^{-{L}/{4z_0^2}}\big)}}\left[\phi(z_1,t)\varphi(z_2,t)+\varphi(z_1,t)\phi(z_2,t)\right]\\
\Phi^-(t)&=&\frac{1}{\sqrt{2\big(1-e^{-{L}/{4z_0^2}}\big)}}\left[\phi(z_1,t)\varphi(z_2,t)-\varphi(z_1,t)\phi(z_2,t)\right].
\end{array}
\right.
\end{eqnarray}
Meanwhile, the internal spin states of two-ion system can be expressed by the basis of the triplet states $\left|\downarrow\downarrow\right\rangle,\ \left|\uparrow\uparrow\right\rangle,\ \left|+\right\rangle\equiv\frac{1}{\sqrt{2}}(\left|\uparrow\downarrow\right\rangle+\left|\downarrow\uparrow\right\rangle)$ and the singlet state $\left|-\right\rangle\equiv\frac{1}{\sqrt{2}}(\left|\uparrow\downarrow\right\rangle-\left|\downarrow\uparrow\right\rangle)$. The total states of two ions are the product of the spatial states and the internal spin states. For instance, if the two ions are Bosons, the total states needs to be symmetric, which results in the symmetric basis describing the total states of two ions, i.e., $\big\{\left|\Phi^+\right\rangle\otimes\left|\downarrow\downarrow\right\rangle,\ \left|\Phi^+\right\rangle\otimes\left|\uparrow\uparrow\right\rangle,\ \left|\Phi^+\right\rangle\otimes\left|+\right\rangle,\ \left|\Phi^-\right\rangle\otimes\left|-\right\rangle\big\}$.

Now we consider the Coulomb interaction between two ions $V(r)=\frac{Q^2}{4\pi \epsilon_0}\frac{1}{r}$, where $Q$ is the charge of single ion and $r\equiv\sqrt{(x_1-x_2)^2+(y_1-y_2)^2+(z_1-z_2)^2}$ is the distance of two ions in three dimensions. In the quasi-one dimensional condition $\omega_\bot\gg 1$, the transverse motions of two ions are frozen in the ground states. Therefore, we average the Coulomb interaction by the ground states in $x$ and $y$ directions and arrive at the effective Coulomb interaction of
two ion in $z$ direction
\begin{eqnarray}\label{Veff}
V_{\mathrm{eff}}(r_z)=\frac{Q^2}{4\pi \epsilon_0z_0}\sqrt{\frac{\pi\omega_{\bot}}{2}}\exp\left(\frac{\omega_{\bot}}{2}\frac{r_z^2}{z^2_0}\right)
\mathrm{erfc\left(\sqrt{\frac{\omega_{\bot}}{2}}\frac{r_z}{z_0}\right)},
\end{eqnarray}
where $r_z\equiv|z_1-z_2|$ is the distance of two ions in $z$ direction and $\mathrm{erfc}[\bullet]$ is the complementary error function. It can be shown that $V_{\mathrm{eff}}(r_z)\rightarrow \frac{Q^2}{4\pi \epsilon_0 r_z}$ for $r_z\gg z_0\sqrt{2/\omega_{\bot}}$, which recovers the Coulomb's law in one dimension.
In the left figure of Fig.~\ref{fig_CIT}d, we show the internal level structure of two ions in the free space without Coulomb interaction. In this case, the two internal states $\left|+\right\rangle$ and $\left|-\right\rangle$ are degenerate. When the two ions are colliding each other in the trap, the Coulomb interaction modifies the internal energy levels of two ions as illustrated by the right figure in the same figure. When the kinetic energy of two ions is much larger than the Coulomb interaction potential, we calculate the level shifts of states $\left|+\right\rangle$ and $\left|-\right\rangle$ by
\begin{eqnarray}\label{vpn}
\left\{
\begin{array}{lll}
V_+(L)&=&\frac{\omega_z}{2\pi}\int_0^{\frac{2\pi}{\omega_z}}\left\langle\Phi^+(t)\right|V_{\mathrm{eff}}(z_1-z_2)\left|\Phi^+(t)\right\rangle dt\\
V_-(L)&=&\frac{\omega_z}{2\pi}\int_0^{\frac{2\pi}{\omega_z}}\left\langle\Phi^-(t)\right|V_{\mathrm{eff}}(z_1-z_2)\left|\Phi^-(t)\right\rangle dt
\end{array}
\right.
\end{eqnarray}
As a result, the degeneracy of states $\left|+\right\rangle$ and $\left|-\right\rangle$ is lifted by a splitting $2J$ with $J\equiv \frac{1}{2}(V_+-V_-)$ as labelled in the right figure of Fig.~\ref{fig_CIT}d, which is similar to the effective spin-spin interaction in Heisenberg model rooting from the exchange interaction of two neighbouring electrons. There is also a common level shift of all internal levels $U\equiv \frac{1}{2}(V_++V_-)$, which is just the time-averaged direct Coulomb interaction.

We discuss how to generate the universal two-qubit $\sqrt{SWAP}$ gate using the collision-induced spin-spin interaction $J$.
From the relationships $\left|\downarrow\uparrow\right\rangle=\frac{1}{\sqrt{2}}(\left|+\right\rangle-\left|-\right\rangle)$ and $\left|\uparrow\downarrow\right\rangle=\frac{1}{\sqrt{2}}(\left|+\right\rangle+\left|-\right\rangle)$, the time evolutions of initial states $\left|\downarrow\uparrow\right\rangle$ and $\left|\uparrow\downarrow\right\rangle$ are given by $\frac{1}{\sqrt{2}}( e^{i\frac{E_0+U+J}{\hbar}t}\left|+\right\rangle-e^{i\frac{E_0+U-J}{\hbar}t}\left|-\right\rangle)$ and $\frac{1}{\sqrt{2}}( e^{i\frac{E_0+U+J}{\hbar}t}\left|+\right\rangle+e^{i\frac{E_0+U-J}{\hbar}t}\left|-\right\rangle)$ respectively. The time evolutions of states $\left|\downarrow\downarrow\right\rangle$ and $\left|\uparrow\uparrow\right\rangle$ are given by $e^{i\frac{U+J}{\hbar}t}\left|\downarrow\downarrow\right\rangle$ and $e^{i\frac{2E_0+U+J}{\hbar}t}\left|\uparrow\uparrow\right\rangle$ respectively. After collision time $t_g=\frac{3\pi\hbar}{4 J}$,
the four two-qubit basis $\left|\downarrow\downarrow\right\rangle$, $\left|\downarrow\uparrow\right\rangle$, $\left|\uparrow\downarrow\right\rangle$ and $\left|\uparrow\uparrow\right\rangle$ arrive at
\begin{eqnarray}\label{}
\left|\downarrow\downarrow\right\rangle&\mapsto&e^{i\frac{(U+J)}{\hbar}t_g}\left|\downarrow\downarrow\right\rangle\nonumber\\
\left|\downarrow\uparrow\right\rangle&\mapsto& e^{i\frac{ (E_0+U+J)}{\hbar}t_g}\frac{1}{2}[(1+i)\left|\downarrow\uparrow\right\rangle+(1-i)\left|\uparrow\downarrow\right\rangle]\nonumber\\
\left|\uparrow\downarrow\right\rangle&\mapsto& e^{i\frac{ (E_0+U+J)}{\hbar}t_g}\frac{1}{2}[(1-i)\left|\downarrow\uparrow\right\rangle+(1+i)\left|\uparrow\downarrow\right\rangle]\nonumber\\
\left|\uparrow\uparrow\right\rangle&\mapsto&e^{i\frac{ (2E_0+U+J)}{\hbar}t_g}\left|\uparrow\uparrow\right\rangle.\nonumber
\end{eqnarray}
This unitary time evolution defines a two-qubit quantum gate in the basis of $\big\{\left|\downarrow\downarrow\right\rangle,\left|\downarrow\uparrow\right\rangle,\left|\uparrow\downarrow\right\rangle,\left|\uparrow\uparrow\right\rangle\big\}$
\begin{eqnarray}\label{Qgates}
U(t_g)\equiv
e^{i\frac{ (E_0+U+J)}{\hbar}t_g}\left(
\begin{array}{llll}
e^{-i\frac{ E_0}{\hbar}t_g}&0&0&0\\
0&\frac{1}{2}(1+i)&\frac{1}{2}(1-i)&0\\
0&\frac{1}{2}(1-i)&\frac{1}{2}(1+i)&0\\
0&0&0&e^{i\frac{ E_0}{\hbar}t_g}\\
\end{array}
\right).\ \ \
\end{eqnarray}
The two-qubit $U(t_g)$ gate can be decomposed into the standard two-qubit $\sqrt{SWAP}$ gate \cite{PRA-57-120-1998,PRA-67-032301-2003} and two single-qubit phase shift gates, i.e., $U(t_g)=\sqrt{SWAP}\left(R_\theta\otimes R_\theta\right)$, where the single-qubit phase gate is defined by $R_\theta\equiv\mathrm{diag}\left(1,\theta\right)$ with phase parameter $\theta={t_g (2E_0+U+J)}/{2\hbar}$.
The two-qubit $\sqrt{SWAP}$ gate between more remote ions, e.g., the ones at two ends of the architecture shown in Fig.~\ref{fig_CIT}a, can be realized by swapping the states of their neighboring ions sequentially using two-qubit SWAP gate.
As the two-qubit $\sqrt{SWAP}$ gate is universal \cite{2010-Nielsen,PRA-57-120-1998,PRA-67-032301-2003}, hence the two-qubit $U(t_g)$ gate introduced here is also universal meaning any quantum algorithm can be realized by the universal two-qubit gate plus single-qubit gates. Therefore, our proposed ion trap architecture is capable of realizing universal quantum computation.

The collision-induced spin-spin interaction determines the speed of two-qubit gate. In the long-distance limit $L\gg z_0$, the analytical expression of $J$ can be obtained directly from Eq.~(\ref{vpn}), i.e.,
\begin{eqnarray}\label{UJ}
J\sim\frac{Q^2}{4\pi\epsilon_0z_0}\frac{\omega_\bot}{\sqrt{\pi}}\left(\frac{z_0 }{
L}\right)^3 A(L/z_0,\omega_\bot)
\end{eqnarray}
with the interference term $A(L/z_0,\omega_\bot)$ given by
\begin{eqnarray}\label{AI}
A\equiv\frac{2L^2}{\omega_\bot z_0^2}\int_{0}^{\infty} e^{\frac{\omega_\bot-1}{2}(\frac{z}{z_0})^2}\mathrm{erfc}
\big(\sqrt{\frac{\omega_\bot}{2}}\frac{z}{z_0}\big)\cos\big[\frac{2p(t)}{\hbar}z\big]dz.
\end{eqnarray}
Here, $2p(t)=m\omega_zL\sin(\omega_zt)$ is the relative momentum of two ions. The main contribution of $A(L/z_0,\omega_\bot)$ comes from the quantum exchange effect during collision. Due to the repulsive Coulomb interaction, the two ions need enough kinetic energy to overcome the barrier of the effective Coulomb interaction potential $V_{\mathrm{eff}}(0)=\frac{Q^2}{4\pi \epsilon_0z_0}\sqrt{\frac{\pi\omega_{\bot}}{2}}$, i.e., $\alpha\equiv\frac{1}{4}m\omega_z^2L^2 \left(\frac{Q^2}{4\pi \epsilon_0z_0}\sqrt{\frac{\pi\omega_{\bot}}{2}}\right)^{-1}>1.$
In the long-distance limit $L\gg z_0\sqrt{\frac{\omega_\bot}{2}}$, we find that $A(L/z_0,\omega_\bot)\rightarrow\frac{2}{\sqrt{\pi}}$ and
\begin{eqnarray}\label{JLD}
J\sim \frac{Q^2}{4\pi\epsilon_0z_0}\frac{2\omega_\bot}{\pi}\left(\frac{z_0 }{L}\right)^3=\frac{Q^2}{2\pi^2\epsilon_0}\frac{\hbar\omega_{x,y}}{m\omega^2_z}\frac{1}{L^3}.
\end{eqnarray}
We see that the collision-induced effective spin-spin interaction $J$ decreases with the third power of the trapping distance $L$, which coincides with the behavior of real spin-spin interaction through magnetic dipole-dipole coupling \cite{MES-1998}.

We discuss the possibilities and challenges in the state-of-art ion trap experiments.
In general, we need relatively large $J$ for high-speed two-qubit gate operation.  For a fixed parameter $\alpha>1$, we have the alternative expression of Eq.~(\ref{JLD})
\begin{eqnarray}\label{JLD2}
J\sim \left(\frac{\hbar}{2\pi}\right)^{\frac{3}{2}}\sqrt{\frac{\omega_{x,y}}{m}}\ \frac{2}{\alpha}\ \frac{1}{L}.
\end{eqnarray}
We can maximize the value of $J$ by taking $\alpha=1$. In this case, we have $\omega_\bot=\Big(\frac{Q^2}{4\pi\epsilon_0}\Big)^{-\frac{1}{2}}\Big(\frac{m\hbar\omega_{x,y}^3}{8\pi}\Big)^{\frac{1}{4}}L$.
We see that $J$ is proportional to the inverse of trapping distance $L$ while $\omega_\bot$ is linear to $L$. In total, we have the restriction condition $\omega_\bot J=\Big(\frac{Q^2}{8\pi\epsilon_0}\Big)^{-\frac{1}{2}}\Big(\frac{\hbar}{2\pi}\Big)^{\frac{7}{4}}\Big(\frac{\omega_{x,y}^5}{ m}\Big)^{\frac{1}{4}}$. In principle, we can increase the transverse trapping frequency $\omega_{x,y}$ to have arbitrary large $J$. However, it is challengeable to make the  trapping frequency $\omega_{x,y}$ beyond $2\pi\times \SI{10}{\mega\hertz}$ in the current ion trap experiments. At the same time, we also need strong transverse confinement $\omega_\bot\gg 1$ to avoid exciting the transverse motion.
By setting the parameter $\omega_{x,y}=2\pi\times \SI{10}{\mega\hertz}$, we have $\omega_\bot J/\hbar\approx \SI{0.94}{\kilo\hertz}$ for the ion $^{171}\mathrm{Yb}^+$ and $\omega_\bot J/\hbar\approx \SI{1.97}{\kilo\hertz}$ for the ion $^{9}\mathrm{Be}^+$. Allowing $\omega_\bot=5$, we have $J/\hbar\approx \SI{0.19}{\kilo\hertz}$ for the ion $^{171}\mathrm{Yb}^+$ with trapping distance $L\approx\SI{103}{\micro\meter}$ and $J/\hbar\approx \SI{0.39}{\kilo\hertz}$ for the ion $^{9}\mathrm{Be}^+$ with with trapping distance $L\approx\SI{215}{\micro\meter}$.
From Eq.~(\ref{JLD2}), we see that it is also possible to obtain large $J$ by choosing particle with small mass. Therefore, the Penning traps could be a better choice since the mass of electron is four orders smaller than the mass of atomic ion and the transverse confinement frequency $\omega_{x,y}$ can reach up to hundreds of $\SI{}{\giga\hertz}$ \cite{PS-34-47-1991,IJMSIP-88-1989}. For the transverse confinement frequency $\omega_{x,y}=2\pi\times\SI{100}{\giga\hertz}$ and the typical trapping distance $L=\SI{10}{\milli\meter}$, we have $J/\hbar\approx \SI{0.108}{}\alpha^{-1}\SI{}{\mega\hertz}$. In the case of $\alpha=1$, we have the spin-spin interaction $J/\hbar\approx \SI{0.108}{\mega\hertz}$ with transverse confinement $\omega_\bot\approx 2.05\times 10^4$.
For the parameters discussed above, the two-qubit gate tim $t_g=\frac{3\pi\hbar}{4J}$ is much shorter than the time period of oscillations. The two ions need time $\Delta t=\pi/\omega_z$ to collide once and exchange their positions. Thus, the number of collisions is determined by $N_g\equiv\frac{t_g}{\Delta t}=\frac{3\hbar\omega_z}{4J}$, which can be finely tuned to be an integer by adjusting experimental parameters.
Our proposed ion trap architecture for universal computation has scalability advantage compared to other proposals such as coupling ions by phonons of Coulomb crystal \cite{PRL-74-1995,PRL-82-1999,PRL-91-2003}. Different from confining a chain of ions in a single trap, we only confine one ion in each trap as sketched in Fig.~\ref{fig_CIT}a. Therefore, arbitrary number of ions can be coupled without the stability problem. Instead,
the technical challenge to realize our proposal comes from the precise time sequential control of the electric potentials on the electrodes. For instance, in the process shown in the right figure of Fig.~\ref{fig_CIT}b, it is optimized to shut off the collision when the two ions just arrive at the bottoms of two separated traps.

We discuss the influences of the direct interaction $U$ on the gate operations. In the long-distance limit $L\gg z_0$,
we calculate the direct interaction from Eq.~(\ref{vpn}), i.e.,
$U\sim\frac{\tilde{Q}^2}{4\pi\epsilon_0L},$
where $\tilde{Q}\equiv Q\sqrt{2\pi^{-1}\ln(2^{3/2}e^{\gamma/2}\omega_\bot^{1/2}{L}/{z_0})}$ is the renormalized charge with the Euler-Mascheroni constant $\gamma=0.5772\cdots$.
In our proposal, the two-qubit gate is realized by oscillating the two ions in the traps. Due to the Coulomb interaction in the colliding process, there is a position shift from the initial positions of two ions after collisions, i.e., the two ions shown in the right figure of Fig.~\ref{fig_CIT}c can not arrive at the bottoms of two traps exactly. This position shift can be accumulated after many times of collisions, which could introduce significant errors to the two-qubit gate operations. To suppress this kind of error accumulation, one possible solution is to add a small parametric driving to the harmonic potential, i.e., the single-ion Hamiltonian is given by $H_0(t)=\frac{p^2}{2m}+\frac{1}{2}m\omega_z^2[1+f\cos(\omega_ft)]z^2-\frac{1}{4}\zeta z^4$. Here, parameters $\omega_f$ and $f$ are the frequency and amplitude of the parametrical driving respectively. Since the real confinement potential is not perfectly harmonic, we also introduce a small nonlinearity $\zeta$ to the confining potential. In the resonant condition $\omega_f=2\omega_z$, the ground state of the parametric oscillator is unstable and two oscillating states with opposite phase become stable, which is the phenomenon of bistability \cite{PRA-042108-2006}. From the standard stability analysis of classical parametric oscillator, the amplitude of the stable oscillating state is given by $\sqrt{{2fm\omega_z^2}/{3\zeta}}$. To reach optimal control, the two ions are needed to be tuned carefully just occupying the bistable states in the experiment, i.e., $L=2\sqrt{{2fm\omega_z^2}/{3\zeta}}$ or $f=\frac{3}{8}\zeta L^2/m\omega_z^2$. The maximum position shift of two ions from the bottoms can be estimated linearly by the theory of phase space interaction \cite{arXiv-1710-09716} $\frac{\Delta L}{L}\sim\frac{4}{fm\omega_z^2L}\frac{dU}{dL}\approx \frac{4}{fm\omega_z^2L^2}U(L)$. For the ion $^{171}\mathrm{Yb}^+$ with parameters discussed above, we have the condition ${\Delta L}/{L}\sim 1.4\times 10^{-4}f^{-1}\ll 1$ setting a lower boundary for the driving power, i.e., $f\gg 1.4\times 10^{-4}$. We can use the external parametric driving field to count the collision number $N_g=\frac{3\hbar\omega_f}{8J}$ and improve the time sequential control of gate operations. The parametric driving can also compensate the energy loss due to the damping in the trap and the radiation of charge oscillations.

Finally, we justify the approximation used in Eq.~(\ref{vpn}) where we have assumed that the perturbation of Coulomb interaction on the harmonic oscillations of two ions is small. From Eq.~(\ref{Veff}), the characteristic length of interaction potential is $z_0\sqrt{2/\omega_\bot}$. In the region of $r_z< z_0\sqrt{2/\omega_\bot}$, the interaction potential can be estimated by $V_{\mathrm{eff}}(r_z)\approx \frac{Q^2}{4\pi \epsilon_0z_0}\sqrt{\frac{\pi\omega_{\bot}}{2}}\left(1-\sqrt{\frac{2\omega_{\bot}}{\pi}}\frac{r_z}{z_0}\right)$. In the case of $\alpha=1$, the velocity of two ions is given approximately by $v\approx \sqrt{\frac{Q^2\omega_\bot r_z}{4\pi m\epsilon_0z^2_0}}$. Thus, the interacting time of two ions can be estimated by $\Delta\tau\sim \int_0^{z_0\sqrt{2/\omega_\bot}}\frac{dr_z}{v}=(\frac{Q^2}{4\pi \epsilon_0})^{-1/2}m^{1/2}z_0^{3/2}(\frac{2}{\omega_\bot})^{3/4}$. In the long trapping distance limit, the interacting time is small compared to the time period of harmonic oscillation, i.e.,  $\Delta\tau\omega_z/2\pi\approx \pi^{-3/4}(\frac{2}{\omega_\bot})^{1/2}(\frac{z_0}{L})\ll 1$ for $L\gg z_0$. In the calculations of $J$ and $U$ above, we have neglect the slowing down effect due to the repulsive Coulomb interaction. To further consider the slowing down of two ions during collision, we should deduct the interaction energy and replace the relative momentum $2p(t)$ in Eq.~(\ref{AI}) by a smaller value, which results in a larger interference term $A(L/z_0,\omega_\bot)$ in Eq.~(\ref{UJ}). Therefore, the measured the collision-induced spin-spin interaction could be larger than the result given by Eq.~(\ref{JLD}).
In the opposite case $\alpha\ll 1$, the two ions cannot touch each other and are merely classical particles.
In this case, there is no exchange interaction ($J=0$) and the direct interaction is given by $U(L)=\frac{\tilde{Q}^2}{4\pi\epsilon_0L}$ with renormalized charge \cite{PRA-053616-2016} $\tilde{Q}\equiv Q\sqrt{2\pi^{-1}\ln({2\pi e^2\epsilon_0 m \omega_z^2L^3}/{Q^2})}$.

In summary, we have proposed a scalable  ion trap architecture for universal quantum computation. The universal two-qubit $\sqrt{SWAP}$ gate
is realized via the collision-induced spin-spin interaction (collision spin effect). We have shown that it is possible to realize two-qubit gate with speed in $\SI{0.1}{\mega\hertz}$ regime using atomic ions with hundreds-of-micrometer trapping distance in Paul traps. Using electrons Penning traps, the speed can be further increased up into $\SI{0.1}{\mega\hertz}$ regime even at centimeter trapping distance .

\bigskip

\textbf{Acknowledgements} We thank Dr. M. Marthaler and Prof. G. Sch\"on for comments on this work. We thank Dr. X. Zhang and Prof. F. Marquardt for very helpful discussions. This work was supported by the Nachwuchsf\"orderprogramm Carl Zeiss Stiftung (0563-2.8/508/2).

%\bigskip
%
%\textbf{Acknowledgements} We thank M. Marthaler and G. Sch\"on for comments on the work. This work was
%supported by the Nachwuchsf\"orderprogramm Carl Zeiss Stiftung (0563-2.8/508/2).
%
%\textbf{Competing interests statement} The authors declare that they have no competing financial
%interests.
%
%\textbf{Correspondence} and requests for materials should be addressed to L.G.
%(e-mail: guolingzhen@hotmail.com).

\end{document}